\documentstyle[12pt]{article}

\title{Canonical quantization of non-local field equations}
\author{D.\ G.\ Barci \\Instituto de F\'\i sica, Universidade Federal
do Rio de Janeiro \\
C.P. 68528, Rio de Janeiro, RJ, 21945-970, Brasil. \and L.\ E.\ Oxman
\\ Departamento de F\'\i sica, Facultad de Ciencias Exactas y
Naturales\\
Universidad de Buenos Aires, Ciudad Universitaria, 1428,\\ Buenos
Aires, Argentina. \and M.\ Rocca \\ Departamento de F\'\i sica,
Universidad Nacional de  La Plata, \\ C.C. 67, (1900) La Plata,
Argentina.}

\sf

\begin{document}
\baselineskip 20pt
\maketitle

\begin{abstract}

We consistently quantize a class of relativistic non-local field
equations characterized by a non-local kinetic term in the
lagrangian. We solve the classical non-local equations of motion for
a scalar field and evaluate the on-shell hamiltonian.
The quantization is realized by imposing Heisenberg's equation which
leads to the commutator algebra obeyed by the Fourier components of
the field. We show that the field operator carries, in general,
a reducible representation of the Poincare group.

We also consider the Gupta-Bleuler quantization of a non-local gauge
field and analyze the propagators and the physical states of the
theory.

\end{abstract}


\newpage

\section{Introduction}

The interest in non-local field theories has always been present in
theoretical physics and it has been associated to many different
motivations.

In Ref. \cite{Feynman-Wheeler}, Wheeler and Feynman
considered these theories as a description of the interaction
between charged particles where the electromagnetic field
does not appear as a dynamical variable. Before the renormalization
theory became well established, physicists considered the possibility
of formulating a finite theory, in order to describe elementary
particle interactions by means of higher order lagrangians or
non-local lagrangians. Pais and Uhlembeck \cite{Pais-Uhlembeck} were
the first in analyzing non-local theories in this context.

More recently, there were efforts to use non-local
theories in connection with the understanding of quark confinement
and anomalies \cite{volkov-efimov}\cite{krasnikov}, and also in
string field theories containing non-local vertices
\cite{eliezer-woodard}\cite{hata}.

Another aspect of non-local theories, is the possibility of relating
them to a regularization scheme. Specifically, the analytic
regularization introduced by Bollini and Giambiagi in Ref.
\cite{anal-reg} can be thought as associated to a non-local kinetic
term in the lagrangian. On the other hand, dimensional regularization
\cite{dim-reg}\cite{TV} does not admit a lagrangian formulation for
non-integer values of the regulating parameter $d$, and although
the Pauli-Villars regularization \cite{pauli-villars} admits a
lagrangian formulation, the corresponding canonical quantization
leads to an indefinite-metric Fock-space and the related unitarity
problem.

Non-local effective theories, containing non-local kinetic terms,
also arise when integrating over some degrees of freedom that belong
to an underlying local field theory (see Refs. \cite{BV}\cite{DD}).

Then, among the different non-local theories that can be formulated,
it is natural to ask for the possibility of a consistent quantization
of theories containing non-local kinetic terms.

At the classical level, Bollini and Giambiagi
\cite{powers-dalambertian} have studied non-local equations
containing arbitrary powers of the d'alambertian operator and in
particular they established a non-trivial relation between
the space-time dimension and the power of the d'alambertian,
in order to satisfy the Huygens principle \cite{dalam1/2}. In
particular, in $(2+1)$ dimensions, the usual wave equation $\Box \phi
= j$ leads to a Green function that does not satisfy the Huygens
principle, while the non-local equation $\Box^{1/2}\phi =j$ does
satisfy this principle.

So, it is not by chance that the pseudo-differential operator
$\Box^{1/2}$ also appears in the context of bosonization in $(2+1)$
dimensions. In Ref. \cite{marino}, Marino has established the
following mappings,
\begin{eqnarray}
i\bar{\psi}\gamma^\mu\partial_\mu\psi &\leftrightarrow&
\Phi^+\Box^{1/2}\Phi \nonumber\\
i\bar{\psi}\gamma^\mu\partial_\mu\psi &\leftrightarrow&
-\frac{1}{4}F^{\mu\nu}\Box^{-1/2}F_{\mu\nu}
+\frac{1}{2}\theta \epsilon^{\mu\nu\rho}A_\mu\partial_\nu A_\rho
+nqt \nonumber
\end{eqnarray}
where $\psi$ is a two-component Dirac spinor, $\varphi$ is a
complex scalar field, $A_\mu$ is a $U(1)$ gauge field, and $nqt$
are non-quadratic terms that can be eliminated in a long wavelength
approximation. In a very interesting paper by Marino
\cite{marino-proyeccion} this lagrangian appears when $(3+1)$D QED is
projected to a physical plane.
The kinetic term of the $(2+1)$ dimensional effective theory is
proportional to $F^{\mu\nu}\Box^{-1/2}F_{\mu\nu}$. In the static
limit, this term reproduces correctly the $1/r$ Coulomb potential
instead of the usual logarithmic behavior of $(2+1)$D QED, this fact
was first noted in Ref. \cite{dalam1/2}.

In Refs. \cite{barcelos-neto}\cite{marino}\cite{marino-proyeccion},
the quantization of higher order and non-local lagrangians has been
realized using the functional integral formalism.

This technique is naturally formulated in the euclidean space. In
order to go back to Minkowski space it is necessary to perform a Wick
rotation. As it was pointed out in many previous works (see Refs.
\cite{bbr1}\cite{bbr2}\cite{bbr3}\cite{bo}\cite{test}), the analytic
properties of these theories may be highly non-trivial. So, it would
be convenient to quantize these theories following canonical
procedures, which are naturally defined in Minkowski space.

In Ref. \cite{amaral-marino}, Amaral and Marino developed the Dirac
quantization for theories containing fractional powers of the
d'alambertian operator. When using this method some difficulties
arise: there is an infinite set of second class constraints and
the Poisson brackets between canonical variables are not well
defined.

In this work we develope the Schwinger quantization for theories
containing a general non-local kinetic term, including the case
analyzed in Ref. \cite{amaral-marino} as a particular case. The
method is based on the observation that the hamiltonian must be the
generator of time translations, that is, we quantize the theory by
imposing Heisenberg's equation. In this way, we do not need to define
infinite momenta which lead to an infinite number of second class
constraints.

Here we will be interested in Lorentz-invariant actions of the form
\begin{equation}
\label{accion}
S=\int d^\nu xd^\nu y~ \varphi(x) V(x-y) \varphi(y)
\end{equation}
Note that the Klein-Gordon equation $\Box \phi +m^2 \phi=0$
projects onto an irreducible representation of the Poincare group
labeled by the mass $m$. However, for non-local fields, the action
(\ref{accion}) do not lead to a Klein-Gordon type equation, and
therefore the on-shell field will carry, in general, a reducible
representation of the Poincare group. This fact will be reflected in
the mass spectrum of the theory.

Changing variables in (\ref{accion}) we can write this action in a
different way
\begin{eqnarray}
S&=& \int \int d^\nu xd^\nu z~  \varphi(x) V(z)  \varphi(x+z)
\nonumber \\
&=&  \int d^\nu x~ \varphi(x) \left\{ \int d^\nu z V(z)
e^{z.\partial}\right\} \varphi(x) \nonumber \\
&=& \int d^\nu x~  \varphi(x) f(\Box) \varphi(x)
\label{f}
\end{eqnarray}
where we used that $V(z)$ is a function of $z^2$ (due to Lorentz
covariance) and therefore its Fourier transform is a function of
$k^2$ only.

If $f(\Box)$ is a polynomial in $\Box$, we are in the case of a
higher order theory. The Shwinger quantization of different models
belonging to this case has been analyzed in Refs.
\cite{bbr1}\cite{bbr2}\cite{bbr3}\cite{bbor}. But, if
$f(\Box)$ cannot be expanded in a finite series, we are in the
case of the non-local theories that we will consider in this paper.

In the higher derivative case, the model is completely defined by
the roots of the polynomial, that is, by giving the set of ``masses''
that participate in the model. Similarly, we will see that, in
non-local theories, the physical content depends on the zeros and
cuts of the function $f$. That is, the choice of the cuts that leads
to a given analytic determination of $f$ is a physical data that must
be fixed ``a priori''; indeed, the mass spectrum will be given by the
singularities of $f^{-1}$.

In section \S 2 we develope the mode expansion for a field obeying a
general non-local equation. In section \S 3 we evaluate the canonical
hamiltonian, we define the vacuum state and evaluate the propagators.

In section \S 4 we quantize the non-local gauge field and we
define the physical states using the Gupta-Bleuler method to
deal with the gauge invariance of the theory. Then we evaluate the
field propagator. Finally, we interpret the mass spectrum and the
form of the propagators in the case of the $(2+1)$D non-local gauge
theory given by ${\cal L}=-\frac{1}{4}F_{\mu\nu}\Box^{-\frac{1}{2}}
F^{\mu\nu}$, introduced in Ref. \cite{marino-proyeccion}.

\section{Mode expansion for a field obeying a non-local equation}

In this section we will obtain the mode expansion for the on-shell
field that corresponds to the quadratic lagrangian
\begin{equation}
{\cal L}=\frac{1}{2}\phi f(\Box) \phi
\label{L}
\end{equation}
where $f(z)$ is a general analytic function having a
cut contained in the negative real axis. For instance, when
$f(z)=(z+m_0^2)^\alpha$ and $\alpha$ is non integer,
the cut is taken along the real axis, running from $-m_0^2$ to
$-\infty$. Note that in the
Fourier-transformed space the operation $\Box$ amounts to a
multiplication by $-k^2$  ($k^2=k_0^2-{\bf k}^2$), implying that the
functions $f(-k^2)$ we are considering have a cut contained in the
space $k^2\geq 0$ of time-like or light-like vectors.

As it occurs with the poles of $f^{-1}$ (isolated singularities) that
they are associated with a possible isolated
massive mode of the field, we will see that the cuts of $f^{-1}$ (a
continuum of singularities) are associated with a continuum of
massive modes that the field can support; then, the determination we
are using for $f(z)$ simply corresponds to a determination of our
physical system.
That is, the cut contained in $k^2\geq 0$, fixes the possible modes
in the continuum to be normal massive modes ($k^2>0$) or light-like
modes ($k^2=0$).
Note that a light-like mode will be present in the continuum if $z=0$
is a brunch point of $f(z)$; this is the case, for
$f(z)=(z+m_0^2)^\alpha$, when $m_0=0$.

In order to obtain the Euler-Lagrange equation for the system
(\ref{L}) we first expand $f$ in a power series: $f(\Box)=\sum_n
a_n\Box^n$, and then apply higher order lagrangian procedures (see
Ref. \cite{higer-lagrangians}):
\begin{eqnarray}
\lefteqn{\sum_l \Box^l\frac{\partial {\cal L}}{\partial\Box^l \phi}=}
\nonumber \\
&&=\sum_l a_l \Box^l \phi=0
\end{eqnarray}
By summing up the series we obtain the non-local equation of motion:
\begin{equation}
f(\Box) \phi = 0~~~.
\label{EM}
\end{equation}
Now, by Fourier transforming the spatial part of $\phi$,
\begin{equation}
\label{fix}
\phi(x)=\int~d {\bf k} \phi_{{\bf k}}(t)e^{-i{\bf k}.{\bf x}}
\end{equation}
we obtain:
\begin{equation}
f\left( \partial_t^2+{\bf k}^2 \right) \phi_{\bf k}(t)=0~~~.
\label{EMt}
\end{equation}
In the simple case of a Klein-Gordon equation, where $f=\Box+m_0^2$,
a general solution to (\ref{EMt}) can be written as
\begin{equation}
\phi_{\bf k}(t)=\frac{1}{2\pi i}\int_{L_+ + L_-} dk_0~
e^{ik_0t}\times \left[\frac{1}{k_0^2-\omega_0^2}\right]
a(k_0,{\bf k})
\label{KG}
\end{equation}
where $\omega_0=\sqrt{{\bf k}^2+m_0^2}$, and $L_+$ (resp. $L_-$) is a
loop sorrounding the pole at $\omega_0$ (resp. $-\omega_0$) in the
positive (resp. negative) sense. The function $a(k_0,{\bf k})$ is
supposed to be analytic in $k_0$ and after the loop integration we
are left with to different functions of ${\bf k}$ ($a(\omega_0,{\bf
k}), a(-\omega_0,{\bf k})$) representing the arbitrarieness in the
initial conditions.

Now, we will construct a general solution to the equation (\ref{EM}).
In order to do so, we proceed by analogy to the Klein-Gordon case. In
eq. (\ref{KG}), the presence of the pole (isolated singularity)
prevents the paths of integration to be deformed to a point. On the
other hand, when we apply the operator $\partial_t^2-\omega_0^2$ in
(\ref{KG}), an additional factor
$-k_0^2+\omega_0^2=-(k_0-\omega_0)(k_0+\omega_0)$ is produced in the
integrand, the poles are canceled, and the loops can be shrinked to a
point, showing that (\ref{KG}) is a solution to the equation
$\left(\partial_t^2-\omega_0^2\right)\phi_{{\bf k}}(t)=0$.

In the general case, according to the physical determination of $f$,
the singularities are contained in $k^2\geq 0$. For a fixed value of
${\bf k}$, the cuts in the $k_0$ variable are contained in the
intervals $(-\infty,-\omega), (+\omega,+\infty)$, where
$\omega=\sqrt{{\bf k}^2}$. For instance, in the case
$f(z)=(z+m_0^2)^\alpha$, the cuts are given by the points in
$(-\infty ,-\omega_0)$,  $(+\omega_0 ,+\infty)$,
$\omega_0=\sqrt{{\bf k}^2+m_0^2}$.
Then, confronting (\ref{KG}), we propose the general solution to eq.
(\ref{EMt}):
\begin{equation}
\phi_{{\bf k}} (t)=i\int_{\Gamma_+ +\Gamma_-}dk_0 e^{ik_0 t}
\left[ \frac{1}{f(-k^2)}\right]
a(k_0,{\bf k})
\label{fi}
\end{equation}
where $\Gamma_+$ (resp. $\Gamma_-$) is a path surrounding in the
positive (resp. negative) sense, all the
singularities of $1/f(-k^2)$, which are present in the
positive (resp. negative) $k_0$-axis. Actually, when we apply
the operator $f(\partial_t^2+{\bf k}^2)$ on (\ref{fi}), we obtain
\begin{eqnarray}
\lefteqn{f\left( \partial_t^2+{\bf k}^2 \right)
i\int_{\Gamma_+ +\Gamma_-}dk_0 e^{ik_0 t}\frac{1}{f(-k^2)}
a(k_0,{\bf k})=} \nonumber  \\
&&i\int_{\Gamma_+ +\Gamma_-}dk_0 e^{ik_0 t}\frac{f(-k^2) }{f(-k^2)}
a(k_0,{\bf k})=0
\end{eqnarray}
then, while the integration paths in (\ref{fi}) are not
homotopic to zero due to the presence of both types of singularities,
poles and cuts in $f^{-1}$, the application of $f$ turns the
integrand an analytic function of $k_0$.

{}From (\ref{fix}) and (\ref{fi}) we obtain the expression for
the on-shell field associated with equation (\ref{EM}):
\begin{equation}
\phi (x)=i\int_{\Gamma_+ +\Gamma_-}dk e^{ikx}\frac{1}{f(-k^2)} a(k)
\label{fic}
\end{equation}
where $dk$ integrates over the whole of the space-time, with $k_0$
moving along $\Gamma_+ +\Gamma_-$ ($a(k)=a(k_0,{\bf k})$).

Now, we can introduce the distributions $\delta^+ G (k)$ and
$\delta^- G (k)$ ($k_0\in \Re $):
\begin{equation}
\delta^+ G (k)\equiv -i\, \delta \left[ 1/f \right] \theta (k_0)
\makebox[.5in]{,}
\delta^- G (k)\equiv -i\, \delta \left[ 1/f \right] \theta (-k_0)
\label{d+}
\end{equation}
where
\begin{equation}
\delta \left[ 1/f \right] =\left[
\frac{1}{f(-(k_0+i\epsilon)^2+{\bf k}^2)}-
\frac{1}{f(-(k_0-i\epsilon)^2+{\bf k}^2)}\right]
\label{salto}
\end{equation}
so as to obtain
\begin{equation}
\phi (x)=\int dk e^{ikx}[\delta^+ G (k) +\delta^- G (k)] a(k)
\end{equation}
and using $\delta^- G (-k)=\delta^+ G (k)$, results:
\begin{equation}
\phi (x)=\int dk [e^{ikx}a(k)+e^{-ikx}\bar{a}(k)]\delta^+ G (k)
\label{fir}
\end{equation}
where we have also taken into account that $\delta^+ G (k)$ is a real
distribution, and used $a(-k)=\bar{a}(k)$ (when $k_0$ is real) so as
to work with a real $\phi (x)$.

The functional $\delta \left[ 1/f \right]$ is the
``discontinuity functional'' at the cut of $1/f$, and is analogue
to the functional $2\pi i\delta (k^2-m_0^2)$, when
$f(-k^2)=-k^2+m_0^2$ This can be seen by using
$(k^2-m_0^2-i\epsilon)^{-1}=P((k^2-m_0^2)^{-1}) +i\pi \delta
(k^2-m_0^2)$ in eq. (\ref{d+}).

\section{The canonical hamitonian and the field propagator}

If we take the lagrangian
\begin{equation}
{\cal L}=\phi (\sum_{k=0}^na_k\Box^k)\phi
\end{equation}
we can compute the corresponding hamiltonian as the magnitud which is
conserved due to the time-translation symmetry of the system. By
using lagrangian procedures for higher order field theories, the
hamiltonian is obtained from the density (see Ref.
\cite{higer-lagrangians}):
\begin{equation}
T^{00}=\sum_{s,t=0}^n \{ \Box^s \frac{\partial {\cal L}}{\partial
\phi^{(s+t+1)}} \ddot{\phi}^{(t)}-\partial^0 \Box^s
\frac{\partial {\cal L}}{\partial \phi^{(s+t+1)}} \dot{\phi}^{(t)}\}-
{\cal L}
\end{equation}
where $\phi ^{(t)}=\Box^t \phi$. From $\frac{\partial {\cal
L}}{\partial \phi^{(s+t+1)}}=a_{s+t+1}\phi$, it results
\begin{equation}
T^{00}=\sum_{s,t=0}^n a_{s+t+1} (\Box^s \phi \Box^t \ddot{\phi}-
\Box^s \dot{\phi} \Box^t \dot{\phi})-{\cal L}
\end{equation}
Now, if we take ${\cal L}_n=\phi \Box^n \phi$ we have
\begin{equation}
T^{00}_n=\sum_{s+t=n-1}(\Box^s \phi \Box^t \ddot{\phi}-
\Box^s \dot{\phi} \Box^t \dot{\phi})-{\cal L}_n
\end{equation}
where $s$ and $t$ run from $0$ to $n-1$. If we Fourier transform the
field $\phi$ (here we are not using the on-shell field yet), we
obtain:
\begin{equation}
H_n=\int d{\bf x}\int dk dk' e^{ikx}e^{ik'x}\phi (k)\phi (k')
(k_0'^2-k_0 k_0')\sum_{s+t=n-1} (i^2)^{s+t+1}(k^2)^s(k'^2)^t-
\int d{\bf x} {\cal L}_n
\end{equation}
Also we have
\begin{equation}
\sum_{s+t=n-1} (i^2)^{s+t+1}(k^2)^s(k'^2)^t=
(k^2)^{n-1}\sum_{s+t=n-1} (-1)^n(k^2)^{s-n+1}(k'^2)^t=
\frac{(-k^2)^n-(-k'^2)^n}{k^2-k'^2}
\end{equation}
and therefore
\begin{equation}
H_n=\int d{\bf x}\int dk dk' e^{ikx}e^{ik'x}\phi (k)\phi (k')
(k_0'^2-k_0 k_0')\frac{(-k^2)^n-(-k'^2)^n}{k^2-k'^2}-
\int d{\bf x} {\cal L}_n
\label{Hn}
\end{equation}
We will now evaluate the hamiltonian for (\ref{L}). By taking the
development
\begin{equation}
f(z)=\sum_{n=0}^{\infty}a_n z^n
\end{equation}
we have
\begin{equation}
{\cal L}=\phi f(\Box) \phi=\sum _{n=0}^{\infty}a_n \phi \Box^n \phi=
\sum _{n=0}^{\infty}a_n {\cal L}_n
\end{equation}
In this way we get the expression
\begin{eqnarray}
\lefteqn{H=\sum _{n=0}^{\infty}a_n H_n =}\nonumber \\
&&=\int d{\bf x}\int dk dk' e^{ikx}e^{ik'x}\phi (k)\phi (k')
(k_0'^2-k_0 k_0')
\frac{f(-k^2)-f(-k'^2)}{k^2-k'^2}-
\int d{\bf x} {\cal L}\nonumber \\
&&\label{Hoff}
\end{eqnarray}
We can see that in spite of the local validity of the development
we are using for $f$, the final result can be expressed in terms of
$f$, irrespective of the particular coefficients of the series.

Now, let us consider the field on-shell, that is, we use in
(\ref{Hoff}) the expression (\ref{fic}):
\begin{equation}
H=-\int d{\bf x}\int_{\Gamma_+ +\Gamma_-} dk
\int_{\Gamma_+' +\Gamma_-'}dk' e^{ikx}e^{ik'x}
\frac{a(k)a(k')}{f(-k^2)f(-k'^2)}
(k_0'^2-k_0 k_0')\frac{f(-k^2)-f(-k'^2)}{k^2-k'^2}
\end{equation}
Integrating in ${\bf x}$ and then in ${\bf k}'$ we get
\begin{eqnarray}
\lefteqn{H=-\int d{\bf k} \int_{\Gamma_+ +\Gamma_-} dk_0
\int_{\Gamma_+' +\Gamma_-'}dk_0' \frac{e^{i(k_0+k_0')t}}{k_0+k_0'}
a(k_0,{\bf k})a(k_0',-{\bf k})
\frac{k_0'}{f(-{k'_0}^2 +{\bf k}^2 )}}\nonumber \\
&&+\int d{\bf k} \int_{\Gamma_+ +\Gamma_-} dk_0
\int_{\Gamma_+' +\Gamma_-'}dk_0' \frac{e^{i(k_0+k_0')t}}{k_0+k_0'}
a(k_0,{\bf k})a(k_0',-{\bf k})\frac{k_0'}{f(-{k_0}^2+{\bf k}^2)}
\nonumber \\
&& \label{Hi}
\end{eqnarray}
Now, for a given ${\bf k}$, the paths $\Gamma_+$ and
$\Gamma_+'$ (resp. $\Gamma_-$ and $\Gamma_-'$)
surround the points that belong to the cut contained in the interval
$(+\omega, +\infty)$ (resp. $(-\infty, -\omega)$). If in addition we
choose the path $\Gamma_+'$ ($\Gamma_-'$) surrounding
all the
points of the paths $-\Gamma_-$ and $\Gamma_+$ (resp. $-\Gamma_+$ and
$\Gamma_-$), we have:

In the first term of (\ref{Hi}), the integrand has no cuts in the
variable $k_0$ and only presents a pole at $-k_0'$
($a(k_0, {\bf k})$ is analytic in $k_0$). But for the paths we are
considering, we can see that when $k_0'$ belongs to
$\Gamma_+' + \Gamma_-'$, the point $-k_0'$ is not encircled neither
by $\Gamma_+$ nor by $\Gamma_-$. Therefore, making the
$k_0$-integral, we obtain that the first term is equal to zero.

In the second term, the integrand has no cuts in the variable $k_0'$
and only presents a pole at $-k_0$. Again, taking into account the
paths we choosed, we see that for $k_0$ belonging to
$\Gamma_+$, $-k_0$ is encircled by $\Gamma_-'$, while for
$k_0$ belonging to $\Gamma_-$, $-k_0$ is encircled by $\Gamma_+'$.
Then, making the $k_0'$-integral first and using Cauchy's theorem,
results:
\begin{eqnarray}
H&=&2\pi i\int d{\bf k} \int_{\Gamma_+} dk_0
a(k_0,{\bf k})a(-k_0,-{\bf k})\frac{k_0}{f(-{k_0}^2+{\bf k}^2)}
\nonumber \\
&&-2\pi i\int d{\bf k} \int_{\Gamma_-} dk_0
a(k_0,{\bf k})a(-k_0,-{\bf k})\frac{k_0}{f(-{k_0}^2+{\bf k}^2)}
\label{Hf}
\end{eqnarray}
which is time independent as it was expected.

Changing variables $k_0 \rightarrow -k_0$ in the second integral
and recalling that $\Gamma_+$ ($\Gamma_-$) goes in the positive
(negative) sense, we get:
\begin{eqnarray}
H&=&2\pi i\int d{\bf k} \int_{\Gamma_+} dk_0
[a(k_0,{\bf k})a(-k_0,-{\bf k})+a(-k_0,-{\bf k})a(k_0,{\bf k})]
\frac{k_0}{f(-{k_0}^2+{\bf k}^2)}\nonumber
\end{eqnarray}
or in terms of the distribution $\delta^+ G$:
\begin{equation}
H=2\pi \int dk\, k_0\, \delta^+ G (k)
[a(k_0,{\bf k})\bar{a}(k_0,{\bf k})+\bar{a}(k_0,{\bf k})
a(k_0,{\bf k})]
\label{Hr}
\end{equation}
We can see that the hamiltonian $H$ is real ($\delta^+ G$ is real);
we note also, from (\ref{d+}) and (\ref{salto}) that the weight
$\delta^+ G$ is essentially a function of $k^2$:
\begin{equation}
\delta^+ G (k)\equiv i \left[ \frac{1}{f(-k^2 +i\epsilon)}
-\frac{1}{f(-k^2 -i\epsilon)}\right] \theta (k_0)
\label{weight}
\end{equation}
Then, from (\ref{fir}), (\ref{Hr}) and (\ref{weight}), the modes
which are present in the spectrum of the theory are those having
$k^2$ at the cut, where the function $f^{-1}$ has a jump. These modes
participate with a weight $\delta^+ G$. We will denote
the mass spectrum of the model, or equivalently the support of
$\delta^+ G$, by $\cal S$.

Now we will obtain the quantum version for the system (\ref{L}).
Recalling that $H$ is conserved due to the time-translation symmetry
of the system, in order to quantize the theory we require, after
the replacement $a\rightarrow \hat{a}$, $\bar{a}\rightarrow
\hat{a}^{\dagger}$, that the hamiltonian be the generator of
time-translations, that is, we impose Heisenberg's equation:
\begin{equation}
[\phi,H]=i\dot{\phi}
\end{equation}
Using (\ref{fir}) it results as usual:
\begin{equation}
[a,H]=k_0 a
\makebox[.5in]{,}
[a^{\dagger},H]=-k_0 a^{\dagger}
\label{sb}
\end{equation}
then, for $k_0>0$,  $a(k)$  (resp. $a^{\dagger}(k)$) is a lowering
operator (resp. raising operator). The vacuum is the
Poincare-invariant state and is given by:
\begin{equation}
a(k)|0\rangle = 0
\makebox[.5in]{,}
k_0>0
\end{equation}
where $k^2$ belongs to $\cal S$, the support of $\delta^+ G$. Then,
by taking normal ordering, we set the vacuum energy to zero, and
the hamiltonian results (up to a factor)
\begin{equation}
H=\int dk\, k_0\, \delta^+ G (k)
a^{\dagger}(k_0,{\bf k})a(k_0,{\bf k})
\label{Hrq}
\end{equation}
Using now (\ref{fir}), (\ref{Hrq}) and (\ref{sb}) we obtain the
algebra ($k_0>0$, $k'_0>0$):
\begin{equation}
\delta^+ G(k) [a(k),a^{\dagger}(k')]=\delta (k-k')
\makebox[.5in]{,}
[a(k),a(k')]=0
\label{algebra}
\end{equation}
which is valid for $k^2, k'^2 \in {\cal S}$.

The two-point correlation function is:
\begin{eqnarray}
\lefteqn{\langle 0|\phi (x)\phi (y)|0\rangle = \int dk dk'
\delta^+ G(k) \delta^+ G(k')\times}\nonumber \\
&&\langle 0|(e^{ikx}a(k)+e^{-ikx}a^{\dagger}(k))
(e^{ik'y}a(k')+e^{-ik'y}a^{\dagger}(k'))|0\rangle \nonumber \\
&&=\int dk dk'e^{ikx-ik'y}\delta^+ G(k) \delta^+ G(k')
\langle 0|a(k)a^{\dagger}(k')|0\rangle \nonumber \\
&&= \int dk e^{ik(x-y)}\delta^+ G(k)
\label{corre}
\end{eqnarray}
which can also be written as:
\begin{equation}
\langle 0|\phi (x)\phi (y)|0\rangle =\Delta_+(x-y)
\makebox[.5in]{,}
\Delta_+(x)=i \int_{\Gamma_+} dk \frac{e^{ikx}}{f(-k^2)}
\label{D+}
\end{equation}
We clearly see from (\ref{D+}) that $\Delta_+(x)$ is a solution to
the homogeneus equation:
\begin{equation}
f(\Box)\Delta_+(x)=i\int_{\Gamma_+} dk e^{ikx}= 0
\end{equation}
On the other hand, the propagator is:
\begin{eqnarray}
\lefteqn{\langle 0|T\{ \phi (x)\phi (y) \} |0\rangle =
\theta (x_0-y_0) \Delta_+ (x-y) + \theta (y_0-x_0) \Delta_+ (y-x)}
\nonumber \\
&&=\theta (x_0-y_0)\int dk e^{ik(x-y)} \delta^+ G(k)  + \theta
(y_0-x_0)
\int dk e^{ik(x-y)} \delta^- G(k) \nonumber \\
&&=i \int_{\Gamma_F} dk \frac{e^{ik(x-y)}}{f(-k^2)}
\label{F}
\end{eqnarray}
where the path
\begin{equation}
\Gamma_F=\theta (x_0-y_0)\Gamma_+ +\theta (y_0-x_0)\Gamma_-
\end{equation}
can be seen to be equivalent to a path that runs above the cut
which is contained in the interval of negative frequencies
$(-\infty,-\omega)$, and runs bellow the cut contained in the
interval of positive frequencies $(+\omega,+\infty)$.
It is also clear that $F(x)=\langle 0|T\{ \phi (x)\phi (0) \}
|0\rangle$ is $i$ times a Green function:
\begin{equation}
f(\Box)\int_{\Gamma_F} dk \frac{e^{ikx}}{f(-k^2)}=
\int_{\Gamma_F} dk e^{ikx}
\label{eq}
\end{equation}
now, the path $\Gamma_F$ in (\ref{eq}) is equivalent to $R$ (the real
axis) and the integral gives $(2\pi)^4\delta (x)$.
Then, from (\ref{F}), we see that the propagator is the inverse of
the kinetic operator $f(\Box)$ with Feynman's prescription to avoid
the singularities.

As an example we can take, $f(z)=z^{1-\alpha}$, ($0 <\alpha <1$)
determined by the cut ${\cal R}e\, z<0$ so that $f(-k^2)$ has the cut
in the positive $k^2$-axis:
\begin{equation}
{\cal L}=\phi \Box^{1-\alpha} \phi
\label{pow}
\end{equation}
In this case, we have a weight function $\delta^+ G$
given by $2(k^2)_+^{\alpha -1}\sin \pi \alpha \,\theta (k_0)$, where
the distribution $x_+^{\lambda}$ is defined by
\begin{displaymath}
x_+^{\lambda}=\left\{ \begin{array}{ll}
x^{\lambda} &\mbox{if $x>0$} \\
0&\mbox{if $x<0$}
\end{array}  \right.
\end{displaymath}
Note that in this case $\delta^+ G$ is positive definite
($0 <\alpha <1$); then the contribution to the energy of every mode
is positive definite and the resulting quantization does not present
indefinite metric problems (cf. (\ref{algebra})). In the singular
case where
$\alpha =0$, the distribution $(k^2)_+^{\alpha -1}$ has a pole (see
Ref. \cite{gelfand}), the residue is proportional to $\delta (k^2)$
and the weight function results $\delta (k^2)\theta (k_0)=
\frac{1}{2\omega}\delta (k_0-\omega)$, obtaining the usual zero-mass
dispersion relation. Note also that the commutators (\ref{algebra})
which are valid for $k^2, k'^2 \in {\cal S}$, can be written as
$\delta^+ G(k')\delta^+ G(k) [a(k),a^{\dagger}(k')]=\delta^+ G(k')
\delta (k-k')$ (this is what one really gets from Heisenberg's
equation). Then, in the usual Klein-Gordon case, where $\delta^+
G(k)=1/(2\omega)\delta (k_0-\omega)$, we can integrate in $k_0,k_0'$
to obtain the usual commutation relation $[a(\omega, {\bf k}),
a^{\dagger}(\omega', {\bf k}')]=2\omega \delta ({\bf k}-{\bf k}')$.
When we are in a region where $k^2$ belongs to a cut, then $\delta^+
G$ is a well defined non-zero function of $k^2$ and we can write the
commutators according to (\ref{algebra}).

In the general case there is not a unique dispersion relation for
the modes of the field but a continuum of massive modes (those
that belong to the cut), each of which with the corresponding
weight.

\section{The non-local gauge field}

Here we will consider the quantization of a non-local abelian gauge
theory defined by the lagrangian
\begin{equation}
{\cal L}=-\frac{1}{4}F_{\mu\nu} g(\Box) F^{\mu\nu}+{\cal L}_{GF}
\label{G}
\end{equation}
${\cal L}_{GF}$ is the gauge fixing term:
\begin{equation}
{\cal L}_{GF}=\frac{\xi}{2} A_{\mu}g(\Box)\partial^{\mu}
\partial^{\nu} A_{\nu}
\end{equation}
Using the gauge $\xi =1$, we get the Euler-Lagrange equation for
(\ref{G}):
\begin{equation}
\Box g(\Box) A^\mu =0
\label{nme}
\end{equation}
Taking $f(\Box)=\Box g (\Box)$ we obtain the on-shell field (cf.
(\ref{fir}) and (\ref{d+}))
\begin{equation}
A_{\mu} (x)=-i\int dk \sum_{\lambda =0}^3
\epsilon^{(\lambda)}_\mu (k)
[e^{ikx}a^{(\lambda)}(k)+e^{-ikx}\bar{a}^{(\lambda)}(k)]\,
\delta \left[ 1/f \right] \theta (k_0)
\label{gaugef}
\end{equation}
where $\delta [1/f]$ is given by
\begin{equation}
\delta  [1/f] = \delta  \left[ \frac{1}{(-k^2)g(-k^2)}\right]
\end{equation}

The polarization vectors $\epsilon^{(\lambda)} (k)$ are defined, in a
$d+1$-dimensional space-time, according to:
\begin{equation}
\epsilon^{(0)} =n
\makebox[.5in]{,}
(n.n=1~,~ n_0>0)
\end{equation}
\begin{equation}
\epsilon^{(i)}.k =
\epsilon^{(i)}.n=0~,~ \epsilon^{(i)}.\epsilon^{(j)}=-\delta_{ij}
\makebox[.5in]{,}(i=1,...,d-1)
\end{equation}
\begin{equation}
\epsilon^{(d)}=[(k.n)^2-k^2]^{-\frac{1}{2}}(k-(k.n) n)
\end{equation}
where $n$ and $k$ are independent vectors. The equation in
(\ref{nme}) leads at the classical level, taking appropriate boundary
conditions, to the requirement $\partial .A=0$.

At the quantum level we have the algebra ($k_0 >0$):
\begin{equation}
-i\delta [1/f] [a^{(\lambda)}(k),a^{(\lambda') \dagger}(k')]=
-\eta^{\lambda \lambda'}\delta (k-k')
\makebox[.5in]{,}
[a^{(\lambda)}(k),a^{(\lambda')}(k')]=0
\label{algebrag}
\end{equation}
Now, according to the Gupta-Bleuler quantization we impose the gauge
condition by defining the space of physical states that satisfy the
Lorentz condition $\partial .A^-|phys\rangle=0$, where $A^-$ is the
annhilation part of $A$;
this is equivalent to:
\begin{eqnarray}
\lefteqn{L(k)|phys\rangle = 0}\nonumber \\
&&L(k)=(a^{(0)}(k)-\alpha a^{(d)}(k))
\makebox[.5in]{,} \alpha= \frac{\sqrt{(k.n)^2-k^2}}{(k.n)}
\end{eqnarray}
A basis is obtained from the vacuum $|0\rangle$ by repeated
application of $a^{(i) \dagger}(k)$, $i=1,...,d-1$ and
\begin{equation}
M^{\dagger}(k)=a^{(0) \dagger}(k)-\alpha^{-1} a^{(d) \dagger}(k)
\label{Md}
\end{equation}
(these operators commute with $L(k)$).

For $k^2=0$, we have $\alpha =1$. In this case,
$M^{\dagger}=L^{\dagger}$ and then, $L^{\dagger}$ and $L$ commute. A
state containg at least an $M^{\dagger}$-type particle ($k^2=0$) has
zero-norm. If we add such a state to any physical state, the norm of
the latter is not modified.

For $k^2>0$, the operator $L$ annhilates particles having
polarization $k/\sqrt{k^2}$, which is proportional to
$\epsilon^{(0)}(k)+\alpha \epsilon^{(d)}(k)$. The physical massive
states have particles with polarizations $\epsilon^{(i)}$
($i=1,...,d-1$) and the polarization $\epsilon'^{(d)}$, which is
proportional to $\epsilon^{(0)}(k) +\alpha^{-1} \epsilon^{(d)}(k)$,
carried by the
particles created by $M^{\dagger}$.
Here, $\alpha \neq 1$ and $\epsilon'^{(d)}$, $k$ are independent
vectors satisfying $\epsilon'^{(d)}.k=0$.
The physical states containing $M^{\dagger}$-type particles
(none of them having zero mass) have non-zero norm.

If we take a matrix element of the vector field between two physical
states and we add to one of them a state containing a
zero mass $M^{\dagger}$-type particle, then, this matrix element
changes by a gauge transformation; on the other hand, when we add
a massive $M^{\dagger}$-type particle the change is non-trivial.
In this regard, note that in a matrix element of the form
\begin{equation}
g_{\mu}=\langle phys |A_{\mu} M^{\dagger}(k)|\psi \rangle
\end{equation}
only when $k^2=0$ we have $\langle phys |M^{\dagger}(k)=
\langle phys |L^{\dagger}(k)=0$ and we can write
\begin{equation}
g_{\mu}=\langle phys |[A_{\mu},M^{\dagger}(k)]\|\psi \rangle
\end{equation}
to obtain from (\ref{gaugef}), (\ref{Md}) and (\ref{algebrag}):
\begin{equation}
g_{\mu}=\partial_{\mu}\alpha (x)
\makebox[.5in]{,}
\alpha (x)=-(k .n)^{-1}e^{ik .x} \langle phys |\psi \rangle
\end{equation}

Then, the states containing at least a zero-mass $M^{\dagger}$-type
particle are associated to gauge transformations. The coexistence of
massive states and gauge symmetry is possible because of the presence
of the zero mass modes. For instance, if $k^2=0$ does not belong to
the cut of $g$, then $1/f$ has a pole at $k^2=0$; for
$g(\Box)=\Box^{-\alpha}$, ($0<\alpha <1$) we have
$f=\Box^{1-\alpha}$ and $k^2=0$ is a brunch point of $1/f$. In both
cases the on-shell field is a superposition of fields with a given
mass spectrum that contains the zero-mass. Then, the gauge
transformation $A_{\mu}\rightarrow A_{\mu} +\partial_{\mu}\alpha$ can
be thought as operating on the zero-mass sector of the vector field.

We end this section discussing the interpretation of the mass
spectrum and the form of the propagators when we take the $(2+1)$
dimensional gauge theory corresponding to
$g(\Box)=\Box^{-\frac{1}{2}}$:
\begin{equation}
{\cal L}=-\frac{1}{4}F_{\mu\nu}\Box^{-\frac{1}{2}} F^{\mu\nu}+{\cal
L}_{GF} +j^{\mu} A_{\mu}
\label{G'}
\end{equation}
In Ref.
 \cite{marino-proyeccion}, Marino has shown that this lagrangian
describes, at the tree-level, the projection of QED from $(3+1)$ to
$(2+1)$ dimensions. That is, the $(2+1)$D effective action (obtained
by path-integrating over the gauge field)
\begin{equation}
\int dt\, dx\, dy\, j^\mu(x) G^{(2+1)}_{\mu \nu}(x-y)j^\nu (y)
\makebox[.5in]{,}
\Box^{\frac{1}{2}}G^{(2+1)}_{\mu \nu}(x)=\eta_{\mu \nu}(2\pi)^3
\delta (x)
\end{equation}
corresponds to the $(3+1)$D effective action
\begin{equation}
\int d^4 x\,  J^\mu(x) G^{(3+1)}_{\mu \nu}(x-y)J^\nu (y)
\makebox[.5in]{,}
\Box G^{(3+1)}_{\mu \nu}(x)=\eta_{\mu \nu}(2\pi)^4\delta (x)
\end{equation}
when the currents are constrained to live on the plane $z=0$:
\begin{equation}
J^\mu=\delta (z) j^\mu ~~~(\mu=0,1,2)
\makebox[.5in]{,}
J^3=0
\end{equation}
In other words, the projected non-local $(2+1)$D theory given by
(\ref{G'}) displays a static coulombic interaction ($\sim 1/R$)
between charges, instead of the logarithmic behavior present in
$(2+1)$ dimensions, when ${\cal L}=-(1/4)F_{\mu\nu}F^{\mu\nu}$.

This can be seen by noting that, for $z=0$, the Maxwell propagator
(in the Feynman gauge) can be written as:
\begin{eqnarray}
i G^{(3+1)}_{\mu \nu}(x)|_{z=0}&=&i\eta_{\mu \nu}
\int_0^{+\infty} d k_z^2\, (k_z^2)^{-\frac{1}{2}} \int d^3 k
\frac{e^{ik_{(3)}x_{(3)}}}{(-k_{(3)}^2+
k_z^2+i\epsilon)}
\label{pp'} \\
&=&i\pi \eta_{\mu \nu}  \int d^3 k
\frac{e^{ik_{(3)}x_{(3)}}}{(-k_{(3)}^2+i\epsilon)^{\frac{1}{2}}}
\label{pp}
\end{eqnarray}
where $k_{(3)}$ and $x_{(3)}$ are the $(2+1)$D part of the
four-vectors
$k$ and $x$, respectively; and we used $\int_0^{\infty}ds\,
s^{-1/2}(1+\beta s)^{-1}= \beta^{-1/2}B(\frac{1}{2},\frac{1}{2})$.
On the other hand, from (\ref{algebrag}), we obtain the propagator
corresponding to (\ref{G'}), computed as the vacuum expectation value
of the $T$-product for two fields (cf. (\ref{F})):
\begin{equation}
i\eta_{\mu \nu}\int_{\Gamma_F} d^3 k
\frac{e^{ik_{(3)}x_{(3)}}}{(-k_{(3)}^2)^{\frac{1}{2}}}
\end{equation}
which is proportional to the projected propagator given in
(\ref{pp}), as the $i\epsilon$ prescription is equivalent to
integrating over the path $\Gamma_F$  that runs above the cut
$(-\infty,-\omega)$, and runs bellow the cut $(+\omega,+\infty)$.

Then, in the case of the $(2+1)$ dimensional non-local gauge theory
defined by (\ref{G'}), we see that the form of the propagator
coincides with that obtained by projecting QED from $(3+1)$ to
$(2+1)$ dimensions. Also, we can see that the existence of a
continuous mass spectrum in the non-local theory can be traced back
from the dispersion relation $k^2=0$, present in $(3+1)$D QED, which
in the projected theory reads as $k_{(3)}^2=k_z^2$, giving a mass
spectrum that goes from zero to infinity (cf. eq. (\ref{pp'})).

\section{Conclusions}

In this work, we have solved a general class of non-local field
equations characterized by a non-local kinetic term. The obtained
mode expansion for the on-shell field coincides with that proposed in
Ref. \cite{amaral-marino}, where fractional powers of the
d'alambertian operator are considered.

We have seen that a non-local equation projects the field onto a
reducible representation of the Poincare group. That is, the on-shell
field carries a representation which is the direct sum of irreducible
representations labeled by the mass. The possible values of $k^2$
that appear in the representation are the singularities of the
function $f^{-1}(-k^2)$ that characterizes the kinetic term. Each
mode is associated with a corresponding weight $\delta^+ G=\sigma
(k^2)\theta (k_0)$, where $\sigma (k^2)=-i\, \delta \left[ 1/f
\right]$ (cf. eq. (\ref{d+})). Then, we have seen that
the analytic determination of $f(-k^2)$ fixes our physical system,
that is, the mass spectrum of the model. These facts can be displayed
more clearly by using
\[
\delta^+ G=\int ds\, \sigma (s) \delta (k^2-s) \theta (k_0)
\]
in eq. (\ref{fir}), to write the field as
\[
\phi (x)=\int ds\, \sigma (s) \phi_s (x)
\makebox[.5in]{,}
\phi_s (x)=\int dk [e^{ikx}a(k)+e^{-ikx}\bar{a}(k)]\delta (k^2-s)
\theta (k_0)
\]
$\phi_s$ is the expression for the field that corresponds
to particles having mass squared s. Similar expressions can be
obtained for the hamiltonian and propagators, as both quantities are
linear in $\delta^+ G$ (cf. (\ref{Hrq}) and (\ref{corre})).

In order to quantize the theory we have first computed the
hamiltonian and verified that it is conserved when the field is
on-shell. Then, we have imposed Heisenberg's equation and have
obtained the commutation rules obeyed by the Fourier components of
the field.

If the model is characterized by a kinetic term that leads
to a mass spectrum that contains normal modes only ($k^2 \geq 0$) and
$\delta^+ G (k)$ (see eq. (\ref{d+})) has a positive definite sign,
then the energy is positive definite, we can define the vacuum state
in the usual way, and we can construct a Fock-space with a positive
definite metric (cf. eq. (\ref{algebra})). In this case the obtained
propagators are of the Feynman type and it is a simple matter to make
a Wick rotation in order to make contact with the path integral
formulation. This is the case for the theory defined by (\ref{pow})
where the lagrangian ${\cal L}=\phi \Box^{1-\alpha} \phi$ leads to
the mass weight function $2(k^2)_+^{\alpha -1}\sin \pi \alpha
\,\theta (k_0)$.

Applying the canonical formalism to a non-local gauge theory
we obtained a continuum mass spectrum that contains the zero mass
modes. That is, the vector field is a continuum superposition of
modes that has a zero mass component.
The gauge invariance is preserved due to the presence of these
component. When a gauge transformation is performed, we can consider
that the zero mass component of the field changes, while the
rest of the modes remain unchanged. Finally we considered a
particular case, by specializing to $(2+1)$ dimensions and
considering a kinetic term $F_{\mu\nu}\Box^{-1/2}F^{\mu\nu}$. In this
case, we have interpreted the mass spectrum and have shown that the
form of the propagator coincides with that obtained in the context of
the projected effective action of Ref. \cite{marino-proyeccion}.
Essentialy the massive modes of the
electromagnetic field take into account that we are making a model
where the matter is confined to live on a physical plane while the
electromagnetic field is not confined.

\section*{Acknowledgments}
We want to acknowledge Profs. J.\ J.\ Giambiagi, C.\ G.\ Bollini and
E.\ C.\ Marino, for very usefull and stimulating discussions.
This work was partially supported by CNPq, Brasil, and CONICET,
Argentina.

\end{document}